\documentclass[12pt]{article} 
\usepackage{graphicx}
\usepackage{cite}
\usepackage{amssymb}
\usepackage{amsmath}

\usepackage{mathrsfs} 
\newcommand{\eh}{\hfill}\newlength{\sperr}

\newenvironment{proof}{{\settowidth{\sperr}{\bf\rm
Proof}%
\par\addvspace{0.3cm}\noindent\parbox[t]{1.3\sperr}
{\bf\rm P\eh r\eh o\eh o\eh f.\eh }%
}}{\nopagebreak\mbox{}\hfill
$\Box$\par\addvspace{0.3cm}}

\newtheorem{Pa}{Paper}[section]
\newtheorem{Tm}[Pa]{{\bf Theorem}}

\newtheorem{La}[Pa]{{\bf Lemma}}%[section]
\newtheorem{Cy}[Pa]{{\bf Corollary}}%[section]
\newtheorem{Rk}[Pa]{{\bf Remark}}%[section]
\newtheorem{Pb}[Pa]{{\bf Problem}}%[section]

\newcounter{condition}
\renewcommand{\thecondition}{\Roman{condition}}

\newenvironment{cnd}[1]{\par\addvspace{0.3cm}
\begin{sloppypar}\refstepcounter{condition}%
\noindent{\bf #1
\thecondition.}\it{}}{\end{sloppypar}\par\addvspace{0.5cm}}

\def\col{{\rm col \ }}

\def\ov{\overline}
\def\wt{\widetilde}

\def\ve{\varepsilon}
\def\BZ{{\mathbb Z}}
\def\BR{{\mathbb R}}
\def\BC{{\mathbb C}}
\def\s{\sigma}

\def\wh{\widehat}

\def\a{\alpha}

\def\om{\omega}

\def\col{\mathrm{col}\,}

\newcommand{\I}{\mathrm{i}}
\def\BN{{\mathbb N}}

\def\clm{{\mathcal M}}

\def\clv{{\mathcal V}}

\newcommand{\E}{\mathrm{e}}

\def\nn{\nonumber}

\def\const{{\rm const}}

\title{Sliding inverse problems for radial Dirac and Schr\"odinger  equations}
\author{Lev Sakhnovich\footnote{99 Cove ave., Milford, CT, 06461, USA. 
E-mail: lsakhnovich@gmail.com}}
\date{}

\begin{document}
\maketitle

% \textbf{Mathematics Subject Classification (2010):} Primary 35Q20, 82B40;
%Secondary  51K99 \\

\noindent  \textbf{Mathematics Subject Classification:} Primary 34A55, 34E05, 34L40; Secondary 58J51, 33C05.
  
\noindent  \textbf{Keywords.} Sliding inverse problem, half-inverse problem, quantum defect, 
 statistical sum, confluent hypergeometric functions, Coulomb-type potential, asymptotics of solutions.
 
\begin{abstract}

\noindent  
New inverse and half-inverse problems: {\it sliding problems} are introduced.  
In this way several physically important equations are recovered from the quantum defect.
In particular, sliding problems are solved for  radial Schr\"odinger  equation, 
radial  Dirac system and multidimensional Schr\"odinger  equation. 
Systems with Coulomb-type potentials are considered as well.

\end{abstract}

 In this paper we introduce  new inverse and half-inverse problems: {\it sliding
problems}.  In Section \ref{LASec1}
we show that this approach can be used 
for  radial Schr\"odinger  equation, 
radial  Dirac system and multidimensional Schr\"odinger  equation. 
In Section \ref{LASec2} we consider equations with Coulomb-type potentials.

%??? We note that inverse problems for the Dirac-type systems, which are considered in many previous chapters of the book (and are usually called now Dirac systems),
We note that inverse problems for the Dirac-type systems, which are  usually called now Dirac systems,
have    been thoroughly investigated (see, e.g., \cite{CG2, FKRS3, Kre1, la[6], LS, SaA7, SaL50} and references therein).
%??? numerous references in Chapter \ref{DirSyst} and Introduction.
Radial Dirac system was introduced earlier than Dirac-type system and differs from Dirac-type system.

\section{Inverse and half-inverse sliding problems}\label{LASec1}
\subsection{Main definitions and results}
\label{lasub1.1}

{\bf 1.} The radial Schr\"odinger  equation has the form:
\begin{equation} \label{la1.1}  
\frac{ d^{2}y}{ d r^{2}}+\left( {z}-\frac{\ell(\ell+1)}{r^{2}}- q(r)\right) y=0,  \quad  0{<}r<\infty  \quad (q=\ov{q}),
\end{equation}
where $\ell=0, 1, 2, \ldots $ Some recent results and references on this well-known equation see, for instance, in \cite{AHM1, KoSaTe0, KoSaTe1}.
The radial  Dirac system has the form:
 \begin{align}&
 \label{la1.2}  
\left(\frac{d}{ d r}+\frac{\ell }{r}\right) f_1-({z} +m-q(r))f_2=0,  \\ &
 \label{la1.3}  
\left( \frac{ d}{ d r}-\frac{\ell }{r}\right) f_2+({z} -m-q(r))f_1=0, 
\\ & \label{la1.3'}
0{<}r<\infty ,  \quad m>0, \quad q=\ov{q}, \quad \ell =\pm 1, \pm 2, \ldots,
\end{align}
where $m$   takes  positive (not necessarily integer) values
and stands for mass.
We easily rewrite \eqref{la1.2} and \eqref{la1.3} in the matrix form
\begin{equation}
\label{la1.9}  
\frac{ d}{ d r}f=(\I z \s_2+V)f,
\end{equation}
$f=\col [f_1, \, f_2]$, whereas the Pauli matrix $\s_2$ and matrix function $V$ have the form:
\begin{equation} \label{la1.10}  
\s_2=\left[\begin{array}{cc}
                     0& -\I \\
                   \I & 0
                  \end{array}\right], \quad V(r)=\left[\begin{array}{cc}
                                             -\ell /r & m-q(r) \\
                                             m+q(r) & \ell /r
                                           \end{array}\right]. 
\end{equation}
The peculiarity of the radial Dirac
system consists
 in the fact that the entry  $\ell/r$ of the matrix $V(r)$
is known  and $(\ell/r){\notin}L^1(0,\, \infty)$. A well-posed inverse problem for the radial Dirac system was absent.
This fact was discussed in our paper \cite{la[11]}. In the present paper we introduce
the notion of the quantum defect $\delta(r)$ at  infinity ($z=\infty$). The corresponding sliding half-inverse problem for Dirac system is well-posed.

Without loss of generality
we consider only the case $\ell >0$ since the equality  \eqref{la1.9} yields the transformed equality, where $Jf$ is substituted for $f$, $-z$ is substituted for
$z$, and  $-q$ and $-\ell $ are substituted for $q$ and $\ell $, respectively (in the expression for $V$). Here $J$ is given by 
$J=\begin{bmatrix} 0 & 1\\ 1 & 0\end{bmatrix}.$

Using quantum defect, we solve first a sliding inverse problem
for  the case of the radial Schr\"odinger equation.
Consider  equation \eqref{la1.1} with  the  boundary conditions
\begin{equation} \label{la1.4}  y(0)=y(a)=0,
\end{equation} 
and assume that
\begin{equation} \label{la1.5}  \int_{0}^{a}|q(t)| d t<\infty , \quad  0<a<\infty .
\end{equation}
The corresponding eigenvalues (i.e., the values of $z$ such that the solutions $y$ satisfying \eqref{la1.4} exist) are  denoted by ${z}_n(\ell)$ $\,(n\in \BN)$.
Then we have (as a corollary of Theorem \ref{latm2.1}) the asymptotic equality  
\begin{equation} \label{la1.6}  
\sqrt{{z}_{n}(a, \ell)}=\frac{\pi}{a}(n+\ell/2)+\frac{a \delta(a)-\ell(\ell+1)}{2(n+\ell/2)\pi a}
+o\left( n^{-1}\right)  
\end{equation}
where $\delta(a)$, which
 is defined by the formula
\begin{equation} \label{la1.7}  
\delta(a)=\int_{0}^{a}q(t) d t,
\end{equation}
is called the {\it quantum defect} of the equation \eqref{la1.1}.
Now, we formulate a {\it sliding inverse problem}:

\begin{Pb}\label{laprobl1.1}
Recover the potential $q$ from the given quantum defect
$\delta$.
\end{Pb}
It is immediate from \eqref{la1.7} that the solution of Problem \ref{laprobl1.1} is given by the equality
\begin{equation} \label{la1.8}  
q(a)=\left(\frac{ d}{ d a}\delta\right)(a).
\end{equation}

We deal with the radial Dirac  system in a similar way. 
In order to formulate our result, we introduce a parameter $\ve(z)$, which for the
case of the Dirac system is given by $\ve=\sqrt{m^{2}-{z}^{2}}$.
\begin{Rk}\label{Rkve}
 More precisely, for all ${z}\in \BC$   excluding the cuts  along the  semi-axes  $(-\infty , -m)$ and
 $(m, \infty )$,
we choose a branch of  $\ve=\sqrt{m^2-{z}^2}$  such that $\Re({\ve})>0$ for $z {\ne}0$. 
We introduce $\ve$ on the cuts as the product $\ve= \sqrt{m+{z}} \, \sqrt{m-{z}}$, where
\begin{equation}\nonumber
\arg(\sqrt{m+{z}})=\begin{cases}
0, & \text{if  ${z}>m,$}\\
-\pi/2, & \text{if  ${z}<-m; $}
\end{cases}
\,\,  
\arg(\sqrt{m-{z}})=\begin{cases}
\pi/2, & \text{if  ${z}>m,$}\\
 0, & \text{if  ${z}<-m.$}
\end{cases}
\end{equation}
Hence, we have
\begin{equation} \label{la3.4}    
\arg({\ve})={\pi}/2, \quad {\mathrm{if}}\quad   {z}>m;\quad  \arg({\ve})=-{\pi}/2, \quad {\mathrm{if}}\quad  {z}<-m.
\end{equation}
\end{Rk}
The following theorem is proved in Subsection \ref{lasec3}.
\begin{Tm}\label{latm1.2}
Let the condition
\begin{equation} \label{la1.12}  
\int_{r}^{\infty }|q(t)| d t<\infty , \quad 0<r < \infty 
\end{equation} 
be fulfilled. Then there is a  solution $\wt F(r, \ve, {\ell} )$
of system \eqref{la1.9} $($where $V$ is given by \eqref{la1.10}$)$, such that the relation
\begin{equation} \label{la1.13}  
\wt F (r, \ve , {\ell} ){\sim}-\frac{1}{2\sqrt{m+{z}}}
\E^{-\I (r\ve +\delta (r))}\begin{bmatrix}\I \\ 1\end{bmatrix}, \quad \ve{\to}\infty  ,\quad {z} >m
\end{equation} 
holds and  $\delta(r)$ $($i.e., the quantum defect of this system$)$
has the form
\begin{equation} \label{la1.14}  
\delta(r)=\int_{r}^{\infty }q(t) d t.
\end{equation}
\end{Tm}

So, for Dirac system we  formulate a \textit{sliding half-inverse problem}:

\begin{Pb}\label{laprobl1.3}
Recover the matrix function $V$ in \eqref{la1.9} from the given value ${\ell} $ and quantum defect
$\delta$.
\end{Pb}
Since $V$ has the form \eqref{la1.10}, the solution of  Problem \ref{laprobl1.3}  
is immediate from \eqref{la1.14} or, more precisely, from the equality
\begin{equation} \label{la1.15}  
 q(r)=-\left(\frac{ d}{ d r}\delta\right)(r).
\end{equation}

\begin{Rk}\label{lark1.4}
The notion of half-inverse problem and the first model of such
problem were introduced in the paper \cite{la[8]} (see also \cite{la[9]}).
\end{Rk}

\begin{Rk}\label{lark1.5}
The values of $\delta(r)$ are  spectral characteristics of the radial Dirac
system.
\end{Rk}

\begin{Rk}\label{lark1.8}
Sliding inverse  Problem \ref{laprobl1.1} and half-inverse Problem \ref{laprobl1.3} are well-posed,  that is,  there is a one to one
correspondence between the spectral data $\delta$
 and potentials $q$  (see relations \eqref{la1.7}
 and \eqref{la1.15}, respectively).  Like all the classical inverse problems, these problems are not stable.
 \end{Rk}
 
 \begin{Rk}\label{lark1.9}
 In the process of solving sliding Problems \ref{laprobl1.1} and \ref{laprobl1.3} we use only the spectral data  for positive energy.
 \end{Rk}

{\bf 2.} Next, we consider one-dimensional and multidimensional Schr\"odinger equations.
Putting in \eqref{la1.1} $l=0$, we obtain the one-dimensional Schr\"odinger equation
\begin{equation} \label{la1.16}  
-y^{\prime\prime}+q(x)y={{z}}y, \quad 0{\leq}x{\leq}a, \quad y(0)=y(a)=0.
\end{equation}
\begin{Tm}[\cite{LS}]\label{latm1.6}
Let condition \eqref{la1.5} be fulfilled. Then the eigenvalues ${z}_{n}(a)=z_n$ of
system \eqref{la1.16} satisfy the relation
\begin{equation} \label{la1.17}  
\sqrt{{z}_{n}}=\frac{\pi}{a}n+\frac{1}{2n\pi}\delta(a)
+o(n^{-1} ), 
\end{equation}
where
$\delta(a)$ is given in  \eqref{la1.7}.
\end{Tm}

We note that formulas \eqref{la1.6} and \eqref{la1.17} coincide,  when $\ell=0.$
In Subsection \ref{lasec4}  we prove the assertion:

\begin{Tm}\label{latm1.7}
Let condition \eqref{la1.5} be fulfilled. Then the corresponding statistical sum
\begin{equation} \label{la1.18}  
Z(T)=\sum_{n=1}^{\infty }\E^{-{z}_{n}/T}
\end{equation}
satisfies the relation
\begin{equation} \label{la1.19}  
Z(T)=(a/2)\sqrt{T/\pi}-1/2
+\delta(a)/(2\sqrt{T\pi})+o(1/\sqrt{T}), \quad  T{\to}\infty .
\end{equation}
\end{Tm}
Recall that $T$ stands in physical applications for temperature.
 
 The  $k$-dimensional Schr\"odinger equation has the form
\begin{equation} \label{la1.20}  
-\Delta{y}+q(x)y={{z}}y, \quad x=(x_1, \, x_2, \ldots, \, x_k){\in}G\subset \BR^k.
\end{equation}
We deal here with a model case, where
\begin{align}&\label{als1}
G=G(a_1,a_2, ...)=\{x: \, 0{\leq}x_{i}{\leq}a_{i} \quad (1{\leq}i{\leq}k)\},
\end{align}
 $q$ admits representation
\begin{equation} \label{la1.22}  
q(x)=\sum_{i=1}^{k}q_{i}(x_{i}),
\end{equation}
the functions $q_i$ satisfy inequalities
\begin{equation} \label{la1.23}  
\int_{0}^{a_i}|q_i(x_i)| d x_i<\infty , \quad   0<a_i<\infty , \quad  1{\leq}i{\leq}k,
\end{equation}
 and the boundary condition (on the boundary $\Gamma$  of $G$) is given by
\begin{equation} \label{la1.21}  
y|_{\Gamma}=0 , 
\end{equation}
Now, ${z}_{n, i}$ are the eigenvalues of the problem \eqref{la1.16},   where
$x=x_i$ and $q(x)=q_i(x_i)$, the statistical sum $Z_i$ is given by
\begin{equation} \label{la1.25}  
Z_i(T)=\sum_{n=1}^{\infty }\E^{-{z}_{n, i}/T},
\end{equation}
$\{{z}_{n}\}$ are the
points of the spectrum of problem \eqref{la1.20},  \eqref{la1.21}, whereas
the corresponding statistical sum $Z(T)$ is defined again by the relation \eqref{la1.18}.
Representation \eqref{la1.22} implies that
\begin{equation} \label{la1.26}  
Z(T)=\prod_{i=1}^{k}Z_i(T).
\end{equation}
From Theorem \ref{latm1.7} and equality \eqref{la1.26} we derive the following assertion.
\begin{Tm}\label{latm1.11}
Let the relations \eqref{la1.22} and \eqref{la1.23} hold.  Then
 \begin{equation} \label{la1.27}  
 Z(T)=\left(\frac{T}{4\pi}\right)^{k/2}\left(  \clv_k-\sqrt{\frac{\pi}{T}} \, \clv_{k-1}+
 \frac{\pi}{T}\clv_{k-2}+2\frac{\delta(G)}{T}+o\left(\frac{1}{T}\right)\right) , 
 \end{equation}
 where $T{\to}\infty$ and $\clv_{-1}=0, \quad  \clv_{0}=1 $,
 \begin{equation} \label{la1.28}  
 \clv_k=\prod_{i=1}^{k}a_i, \quad \clv_{k-1}=\clv_k\sum_{i=1}^{k}1/a_i, 
 \quad \clv_{k-2}=\clv_k\sum_{1{\leq}i<j{\leq}k}1/(a_i{a_j}), 
 \end{equation}
 \begin{equation} \label{la1.29}  
 \delta(G)=\int_{G}q(x) d x.
 \end{equation}
 \end{Tm}
 
 The coefficients $\clv_i$ have clear geometric interpretation,  namely, 
 \begin{equation} \label{la1.30}  
 \clv_i=S_i/2^{k-i}, \quad i=k,\,  k-1,\,  k-2, 
 \end{equation}
 where $S_k$ is the Lebesgue measure of the domain $G$,  $S_{k-1}$ is the Lebesgue measure of the boundary $\Gamma$,  $S_{k-2}$ is the Lebesgue measure of the
 domain $\Gamma_1$ formed by the set of pairwise intersections of the facets of the boundary $\Gamma.$ 
 
\noindent 
\textbf{Conjecture I.}\quad 
%\begin{Cre} 
\textit{Formulas \eqref{la1.27}-\eqref{la1.29} hold for a much wider class of potentials than the potentials
of the form \eqref{la1.22}.}
%\end{Cre}

\begin{Pb}\label{laprobl1.12}
Let formulas \eqref{la1.27}-\eqref{la1.29} hold. Recover the potential $q$ from the given $\delta(G)$.
\end{Pb}

According to  \eqref{als1} and \eqref{la1.29} the solution of Problem \ref{laprobl1.12} is given by
\begin{equation} \label{la1.31}  
q(a_{1}, a_{2}, \ldots , a_{k})=
\frac{\partial^{k}}{\partial{a_1}\partial{a_2}\cdots\partial{a_k}}\int_{G(a_{1}, a_{2}, \ldots , a_{k})}q(x) d x.
\end{equation}

{\bf 3.} Finally, we consider one-dimensional and multidimensional  equations for anharmonic oscillators.
The one-dimensional equation for anharmonic oscillator (or one-dimensional  anharmonic oscillator, which is shorter to say)
has the form
\begin{equation} \label{la1.32}  
-y^{\prime\prime}+\left( x^{2}/4+q(x)\right) y={{z}}y, \quad 0{\leq}x<\infty \quad (q=\ov{q}),
\end{equation}
and we introduce the boundary condition
\begin{equation} \label{la1.33}  
y(0)=0.
\end{equation}
We assume that
\begin{equation} \label{la1.34}  
\int_{0}^{\infty }|q(x)|(1+x^{2+\ve}) d x <\infty , \quad  \ve>0. 
\end{equation}

\begin{Tm}[\cite{la[14]}]\label{latm1.13} 
Let condition \eqref{la1.34} be fulfilled. Then the eigenvalues ${z}_{n}$ of
problem \eqref{la1.32}, \eqref{la1.33} satisfy the relation
\begin{equation} \label{la1.35}  
{z}_{n}=2n-1/2+\frac{1}{\pi\sqrt{2n}}(\delta +o(1)), \quad n{\to}\infty  , 
\end{equation}
where
\begin{equation} \label{la1.36}  
\delta=\int_{0}^{\infty }q(x) d x.
\end{equation}
\end{Tm}
Our next theorem easily follows from Theorem \ref{latm1.13}, see Subsection \ref{lasec4}.
\begin{Tm}\label{latm1.14}
Let an anharmonic oscillator satisfy conditions of  Theorem \ref{latm1.13}.
Then the statistical sum $Z(T)$, given by \eqref{la1.18}, satisfies the asymptotic relation
\begin{equation} \label{la1.38}   
Z(T)=T/2-1/4-\delta\left/ \left( 2\sqrt{T\pi}\right)\right. +o\left( 1/\sqrt{T}\right) .
\end{equation}
\end{Tm}
The  $k$-dimensional equation for anharmonic oscillator has the form
\begin{equation} \label{la1.39}  
-\Delta{y}+\left( \frac{|x|^{2}}{4}+q(x)\right) y={{z}}y, \quad x{\in}G,  \quad |x|^{2}=\sum_{i=1}^k{x_i}^{2}.
\end{equation}
We set  
\begin{align}&\label{als2}
G=\{x: \, 0{\leq}x_{i}<\infty \quad (1{\leq}i{\leq}k)\},
\end{align}
and introduce the condition \eqref{la1.21} on the boundary $\Gamma$  of the domain $G$.
We assume again that  the potential $q(x)$ has  the form \eqref{la1.22}
and  that
\begin{equation} \label{la1.42}  
\int_{0}^{\infty }|q_i(x_i)|(1+{x_i}^{2+\ve}) d x_i<\infty , \quad \ve>0 \quad  (1{\leq}i{\leq}k).
\end{equation}
Like in the case of the multidimensional Schr\"odinger equation, ${z}_{n, i}$ are the eigenvalues of the one-dimensional problem, 
this time of the problem \eqref{la1.32},  \eqref{la1.33},   where
$x=x_i$ and $q(x)=q_i(x_i)$.   By ${z}_{n}$
we denote the spectrum of the problem \eqref{la1.39} with initial condition \eqref{la1.21}, where $G$ is given by \eqref{als2}.
The statistical sum $Z_i(T)$ has the form 
\eqref{la1.25} and 
the statistical sum $Z(T)$ has the form
\eqref{la1.18}.  
Like before,   relation \eqref{la1.22} yields \eqref{la1.26}.
From \eqref{la1.26} and \eqref{la1.38} follows the next  assertion.
\begin{Tm}\label{latm1.15} Let a $k$-dimensional anharmonic oscillator, where the potential $q$ admits representation \eqref{la1.22}
and the functions $q_i$ satisfy the condition \eqref{la1.42}, be given.
 Then we have
 \begin{equation} \label{la1.46}  
 Z(T)=\left(\frac{T}{2}\right)^{m}\left( 1-m\left/ (2T)\right. -\delta(G)\left/\left( T\sqrt{T\pi}\right)\right. +o\left( 1\left/ T\sqrt{T}\right. \right)\right) ,  
 \end{equation}
 where $T{\to}\infty$ and $ \delta(G)=\int_{G}q(x) d x$.
\end{Tm}

\noindent 
\textbf{Conjecture II.} \quad 
\textit{Formula \eqref{la1.46} holds for a much wider class of potentials than the potentials
of the form \eqref{la1.22}.} 

\begin{Rk}\label{lark1.16}
Theorem \ref{latm1.15} is an analogue of Theorem \ref{latm1.11}. Jointly with the results from \cite{la[14]},
an easy modification of Theorem \ref{latm1.15}
can be used for the formulation and solution of a  sliding inverse problem for  anharmonic oscillator.
\end{Rk}

\subsection{Radial Schr\"odinger  equation and  quantum defect}
\label{lasec2}
We add variables $\ve=\I \sqrt{{z}}$ and $\ell$ into the notation of  the solutions
of \eqref{la1.1} and write $y(r,\ve,\ell)$.  Here, the parameter $\ve$, continuously depending on $z$, is defined for all ${z}$  in the complex plane
 with the cut along the negative  part of the imaginary  axis. Let us fix $\arg{\ve}={\pi}/2$ for ${z}>0$.
We start our study of  the equation \eqref{la1.1}  from the case $q\equiv 0$.
 Two solutions $u_{1}(r, \ve , \ell )$  and $u_{2}(r, \ve , \ell )$ of  \eqref{la1.1}, where $q\equiv 0$,
 are given by the relations
\begin{equation} \label{la2.2}  
u_{1}(r, \ve, \ell)=(2\ve)^{-(\ell +1)}M_{0, \ell +1/2}(2r\ve ), \quad
u_{2}(r, \ve, \ell)=(2\ve)^{\ell }W_{0, \ell +1/2}(2r\ve ),
\end{equation}
and $M_{\kappa, \mu}(z)$ and $W_{\kappa, \mu}(z)$ in the formula above are  Whittaker functions.
The  connections between  Whittaker functions and
confluent hypergeometric functions are well-known (see, e.g., \cite{la[1]}):
\begin{equation} \label{la2.3}  
M_{0, \ell+1/2}(x)=\E^{-x/2}x^{c/2}\Phi(\a, c, x), \quad
W_{0, \ell+1/2}(x)=\E^{-x/2}x^{c/2}\Psi(\a, c, x), 
\end{equation}
where $\a=\ell +1$,  $c=2\ell +2.$ It follows from \eqref{la2.3}
and relations
\begin{equation} \label{la2.4}  
\Phi(\a, c, x){\sim}1 \quad {\mathrm{and}} \quad \Psi(\a, c, x){\sim}\frac{\Gamma(c-1)}{\Gamma(\a)}x^{1-c} \quad {\mathrm{for}} \quad   x{\to}0,  
\end{equation}
that
the solutions  $u_{1}(r, \ve, \ell)$ and $u_{2}(r, \ve, \ell)$  satisfy the conditions
\begin{equation} \label{la2.5}  
u_1(r, \ve, \ell){\sim}r^{\ell+1}, \quad  
u_2(r, \ve, \ell){\sim}\frac{\Gamma(2\ell+1)}{\Gamma(\ell+1)}r^{-\ell}, \quad  r{\to}0.
\end{equation}
Hence, in view of \eqref{la2.5}  we have
\begin{equation} \label{la2.6}  
|u_{1}(r, \ve, \ell)|{\leq}\clm r^{\ell+1} \quad {\mathrm{for}} \quad  r{\leq}1/|\ve |.
\end{equation}
We note that  the letter $\clm$ stands for different constants.

Let us write the following integral representation (see \cite[Section 6.11]{la[1]}):
\begin{equation} \label{la2.7}  
M_{0, \ell+1/2}(2r\ve)=C(\ell)(2r\ve)^{\ell+1}I(r, \ve, \ell), 
\end{equation}
where
\begin{equation} \label{la2.8}  
C(\ell)=2^{-(2\ell+1)}\frac{\Gamma(2\ell+2)}{\Gamma^{2}(\ell+1)}, \quad 
I(r, \ve, \ell)=\int_{-1}^{1}\E^{rt\ve}(1-t^{2})^{\ell} d t
\end{equation}
Integrating $I(r, \ve, \ell)$  by parts $\ell+2$ times, we  obtain the asymptotic formula:
\begin{equation} \label{la2.9}  
I(r, \ve, \ell){\sim}(r\ve)^{-(\ell+1)}2^{\ell}\Gamma(\ell+1)
\big((-1)^{\ell+1}\E^{-r\ve}\phi(r, \ve, \ell)+\E^{  r\ve}\phi(r, -\ve, \ell)\big), 
\end{equation}
where
\begin{equation} \label{la2.10}  
\phi(r, \ve, \ell)=1+\frac{\ell(\ell+1)}{2r\ve}+o(1/\ve), 
\quad \ve{\to}\infty  .
\end{equation}
Here, we used the Leibnitz differentiation formula in order to  obtain an expression for the $(\ell+1)$'th derivative
of the product $(1-t)^{\ell}(1+t)^{\ell}$.
 Formulas \eqref{la2.2}, \eqref{la2.3} and \eqref{la2.7}-\eqref{la2.9}
imply that
\begin{equation} \label{la2.11}  
u_{1}(r, \ve, \ell)=C_{1}\left( \ell, \ve\right)
\left( -\E^{\I\pi\ell/2}\E^{-r\ve}\phi\left( r, \ve, \ell\right) + \E^{-\I \pi\ell/2}\E^{r\ve}\phi\left( r, -\ve, \ell\right)\right) , 
\end{equation}
where
$C_{1}(\ell, \ve)=\E^{\I \pi\ell/2}\frac{\Gamma(2\ell+2)}{\Gamma(\ell+1)}(2\ve)^{-(\ell+1)}.$

In order to study the asymptotic behavior of $u_{2}(r, \ve, \ell)$ we need
the following integral representation (see \cite[Section 6.11]{la[1]}): 
\begin{equation} \label{la2.12}  
\Gamma(\ell+1)W_{0, \ell+1/2}(x)=\E^{-x/2}x^{\ell+1}
\int_{0}^{\infty }\E^{-tx}t^{\ell}(1+t)^{\ell} d t, \quad x>0.
\end{equation}
Formula \eqref{la2.12} leads us to  the equality
\begin{equation} \label{la2.13}  
\Gamma(\ell+1)W_{0, \ell+1/2}(x)=\E^{-x/2}x^{-\ell}
\int_{0}^{\infty }\E^{-s}s^{\ell}(x+s)^{\ell} d s, \quad x>0.
\end{equation}
In view of the well-known integral representation
\begin{equation} \label{la2.14}  
\Gamma(z)=\int_{0}^{\infty }\E^{-s}s^{z-1} d s,  \quad  \Re{z}>0,
\end{equation} 
equality \eqref{la2.13} takes the form
\begin{equation} \label{la2.15}  
\Gamma(\ell+1)W_{0, \ell+1/2}(x)=\E^{-x/2}x^{-\ell}
\sum_{p=0}^{\ell}C_{\ell}^{p}\Gamma(2\ell-p+1)x^{p}, 
\end{equation}
where $C_{\ell}^{p}$ are the binomial coefficients.  Using analyticity, we rewrite \eqref{la2.15}
for the complex plane
\begin{equation} \label{la2.15'}  
\Gamma(\ell+1)W_{0, \ell+1/2}(z)=\E^{-z/2}z^{-\ell}
\sum_{p=0}^{\ell}C_{\ell}^{p}\Gamma(2\ell-p+1)z^{p},  \quad z\not=0.
\end{equation}
Now,  the asymptotic relation
\begin{equation} \label{la2.16}  
u_{2}(r, \ve, \ell)=(2\ve)^{\ell}\E^{-r\ve}\phi(r, \ve, \ell ), \quad \ve{\to}\infty 
\end{equation}
follows directly from \eqref{la2.2} and \eqref{la2.15'}.

Next we deal with the case  $q{\not\equiv}0$. We assume that $q$ satisfies the condition
\begin{equation} \label{la2.17}  
\int_{0}^{r}|q(t)| d t<\infty , \quad 0<r<\infty  .
\end{equation}
Then  there is a solution $u$ of  \eqref{la1.1} such that the equality
\begin{equation} \label{la2.18}  
u(r, \ve, \ell)=u_{1}(r, \ve, \ell)+\int_{0}^{r}G(r, t, \ve, \ell)q(t)
u(t, \ve, \ell) d t
\end{equation}
holds for $G$ of the form
\begin{equation} \label{la2.19}  
G(r, t, \ve, \ell)=\frac{\Gamma(\ell+1)}{\Gamma(2\ell+2)}\big(u_{1}(r, \ve, \ell)u_{2}(t, \ve, \ell)-
u_{2}(r, \ve, \ell)u_{1}(t, \ve, \ell)\big).
\end{equation} 
Relations \eqref{la2.5} and \eqref{la2.19}  imply that
\begin{equation} \label{la2.20}  
(t/r)^{\ell+1}|G(r, t, \ve, \ell)|{\leq} \clm t, \quad  t{\leq}r{\leq}1/|\ve|.
\end{equation}
From  \eqref{la2.5}, \eqref{la2.18} and \eqref{la2.20} we derive  that the asymptotics of  $u$ is  similar to the asymptotics of  $u_1$, namely
\begin{equation} \label{la2.21}  
u(r, \ve, \ell){\sim}r^{\ell+1}, \quad r{\leq}1/|\ve|.
\end{equation}
Moreover, according to \eqref{la2.10}, \eqref{la2.11}, \eqref{la2.16} and \eqref{la2.19} we have
\begin{equation} \label{la2.22}  
G(r, t, \ve, \ell){\sim}\frac{1}{2\ve}\big(\E^{(r-t)\ve}-\E^{-(r-t)\ve}\big), \quad 
\ve{\to}\infty .
\end{equation}
Using \eqref{la2.5} and \eqref{la2.18}-\eqref{la2.21} we obtain the relation
\begin{equation} \label{la2.23}  
\int_{0}^{1/|\ve|}|G(r, t, \ve, \ell)q(t)u(t, \ve, \ell)| d t=
o\left( |\ve|^{-(\ell+2)}\right) , \quad \ve{\to}\infty .
\end{equation}
It follows from \eqref{la2.22} that for some $\clm$ we have
\begin{equation} \label{la2.24}  
|G(r, t, \ve, \ell)|{\leq}\clm \E^{r|\sigma|}/|\ve|
, \quad 1/|\ve|{\leq}t{\leq}r, 
\end{equation}
where $\sigma:=\Re({\ve})$. In view of \eqref{la2.5}, \eqref{la2.10}, \eqref{la2.11}, \eqref{la2.16} and \eqref{la2.21} the
relation
\begin{equation} \label{la2.25}  
\int_{0}^{1/|\ve|}|G(r, t, \ve, \ell)q(t)
u(t, \ve, \ell)| d t=\E^{r|\sigma|}o(1/|\ve|), \quad r>1/|\ve|, \quad \ve{\to}\infty 
\end{equation}
holds. Formulas \eqref{la2.18}, \eqref{la2.24} and \eqref{la2.25} imply the following equality
\begin{equation} \label{la2.26}  
u(r, \ve, \ell)=u_{1}(r, \ve, \ell)+O\left( \E^{r|\sigma|}/|\ve|\right) , \quad  r>1/|\ve|.
\end{equation}
We denote by ${z}_{n}(a, \ell)$ the eigenvalues of problem \eqref{la1.1}, \eqref{la1.4}. It is well-known that there is only one (up to a constant factor)
solution of \eqref{la1.1} which turns to zero at $r=0$. Since $u$ is such a solution, ${z}_{n}(a, \ell)$ is an  eigenvalue if
and only if
\begin{equation} \label{la2.27}  
u(a, \I \sqrt{{z}_{n}(a, \ell)}, \ell)=0.
\end{equation}
By well-known methods (see, e.g.,  \cite[Ch.1, Section 2]{la[6]}),
using relations \eqref{la2.11}, \eqref{la2.16} and \eqref{la2.26}, \eqref{la2.27}, we obtain  that
\begin{equation} \label{la2.28}  
\sqrt{{z}_{n}(a, \ell)}=\frac{\pi}{a}\left(n+\frac{\ell}{2}\right)+o(1).
\end{equation}
Recall that $\sigma=\Re({\ve})$. Putting
\begin{equation} \nonumber
\Re({\ve})=0, \quad 
(2\ve)^{\ell+1}u(r, \ve, \ell)=\wt u(r, \ve, \ell),  \quad
(2\ve)^{\ell+1}u_1(r, \ve, \ell)=\wt u_1(r, \ve, \ell),
\end{equation}
and
taking into account estimates \eqref{la2.22} and \eqref{la2.26}, we deduce from \eqref{la2.18}   that
\begin{equation} \label{la2.29}  
\wt u(r, \ve, \ell)=\wt u_1(r, \ve, \ell)+
\frac{1}{2\ve}\int_{0}^{r}\left(\E^{(r-t)\ve}-\E^{-(r-t)\ve}\right)\wt u_1(t, \ve, \ell)q(t) d t+o\left(\frac{1}{\ve}\right) .
\end{equation}
Thus, we obtain
 the following assertion.
 \begin{Tm}\label{latm2.1}
Let condition \eqref{la2.17} be fulfilled and equality $\Re(\ve)=0$ hold.  Then the solution $u(r, \ve, \ell)$
of the equation \eqref{la1.1} satisfies \eqref{la2.21}  and admits representation
\begin{equation} \label{la2.30}  
u(r, \ve, \ell) =C_1(\ell, \ve)\left( -\E^{\I\pi\ell/2}\E^{-r\ve+
\Delta(r)/(2\ve)}+\E^{-\I \pi\ell/2}\E^{r\ve-
\Delta(r)/(2\ve)}+o(1/|\ve|)\right), 
\end{equation} 
where $\Delta(r)=\frac{1}{r}\ell(\ell+1)-\delta(r)$
and the quantum defect $\delta(r)$
has the form
\begin{equation} \label{la2.32}  
\delta(r)=\int_{0}^{r}q(t) d t.
\end{equation}
\end{Tm}
Recall that \eqref{la1.6}
follows from Theorem \ref{latm2.1}. 
We denote by $z_{n}(\ell)$ the positive roots of the Whittaker function $M_{0, \ell+1/2}(\I z), $ which are not equal to zero.  
Formula \eqref{la1.6} yields the next well-known result (see \cite[Ch. XI]{la[16]}): 
\begin{equation} \label{la2.34}  
z_{n}(\ell)={\pi}\left(n+\frac{\ell}{2}\right)-\frac{\ell(\ell+1)}{\pi (2n+\ell)}
+o(n^{-1}).
\end{equation}

\subsection{Dirac equation and quantum defect}
\label{lasec3}
{\bf 1.} We assume that $q$ satisfies \eqref{la1.3'} and \eqref{la1.12}, that is,
\begin{equation} \label{la3.1}  
q(r)=\overline{q(r)}, \quad \int_{r}^{\infty }|q(t)| d t<\infty , \quad r>0.
\end{equation}
Given functions $f_1$ and $f_2$, it is easy to find functions $Q_1$ and $Q_2$ such that
\begin{align} \label{la3.2}  &
f_1(r,\ve, \ell)=\sqrt{m+{z}}\, \E^{-r\ve}(2r\ve)^{{\ell} -1}\big(Q_{1}(r,\ve,\ell)+Q_{2}(r,\ve,\ell)\big)r, \\
\label{la3.2'}  &
f_2(r,\ve,\ell)=-\sqrt{m-{z}} \, \E^{-r\ve}(2r\ve)^{{\ell} -1}\big(Q_{1}(r,\ve,\ell)-Q_{2}(r,\ve,\ell)\big)r, 
\end{align}
where $\ve=\sqrt{m^{2}-{z}^{2}}$.
 Recall that  $\ve$ is defined more precisely in Remark \ref{Rkve}.
In order to prove Theorem \ref{latm1.2}, following  \cite{la[2]} we consider solutions of \eqref{la1.2}-\eqref{la1.3'} in the form
\eqref{la3.2} and \eqref{la3.2'}. 
Substituting \eqref{la3.2} and \eqref{la3.2'} into \eqref{la1.2} and \eqref{la1.3} we obtain:
\begin{align} \label{la3.5}& 
 r(Q_{1}+Q_{2})^{\prime}+2{\ell} (Q_{1}+Q_{2})-2r{\ve}Q_{2}+
\sqrt{\frac{m-{z}}{m+{z}}} \,(Q_{1}-Q_{2})rq(r)=0,
\\ & \label{la3.6}  
r(Q_{1}-Q_{2})^{\prime}+2r{\ve}Q_{2}
 -\sqrt{\frac{m+{z}}{m-{z}}}\, (Q_{1}+Q_{2})rq(r)=0,
\end{align}
where $Q^{\prime}=\frac{d}{dr}Q$.     It is easy to see that
\begin{equation}
\label{la3.7} 
 \sqrt{\frac{m-{z}}{m+{z}}}-\sqrt{\frac{m+{z}}{m-{z}}} =
-\frac{2}{\ve}z, \quad \sqrt{\frac{m-{z}}{m+{z}}}+\sqrt{\frac{m+{z}}{m-{z}}} =
\frac{1}{\ve}m.
\end{equation}
Hence, if $q(r)\equiv 0 $,
the
equations \eqref{la3.5}  and \eqref{la3.6} can be rewritten in the form
\begin{equation}
\label{la3.10}
 rQ_{1}^{\prime}+{\ell} Q_{1}+
{\ell} Q_{2}=0,  \quad  rQ_{2}^{\prime}+{\ell} Q_{2} -2r{\ve}Q_{2}+
{\ell} Q_{1}=0. 
\end{equation}
We substitute into \eqref{la3.10} functions 
\begin{align} 
\label{la3.11'} &
Q_i(r,\ve,\ell)=Q_{i,0}\big(\rho/(2\ve),\ve,\ell\big) \quad (i=1,2), \quad \rho = 2r \ve,
\end{align}
and so switch to the new variable  
 $\rho$.
 It follows from \eqref{la3.10} that
\begin{align} 
\label{la3.12} &
\rho \frac{d^2}{d\rho^2}Q_{1, 0}+(2{\ell} +1-\rho)\frac{d}{d\rho}Q_{1, 0}
-{\ell} Q_{1, 0}  =  0, 
\\ & \label{la3.13}  
 {\rho}\frac{d^2}{d\rho^2}Q_{2, 0}+(2{\ell} +1-\rho)\frac{d}{d\rho}Q_{2, 0}
-({\ell} +1)Q_{2, 0}  =  0. 
\end{align}
The 
solutions of \eqref{la3.12} and \eqref{la3.13}, which are regular at the point $\rho=0$, are the confluent hypergeometric functions 
$a_1\Phi({\ell} ,  2{\ell} +1, \rho) $ and $a_2\Phi({\ell} +1,  2{\ell} +1, \rho)$, respectively
(see \cite[Section 36]{la[2]} and \cite[Section 14]{la[3]}). Substitute $r=0$ into \eqref{la3.10}, \eqref{la3.11'} to see that $a_1=-a_2$. That is,
up to the  factor $a_1$ we have 
\begin{align} &
\label{la3.14}  
 Q_{1, 0}(\rho)  =  \Phi({\ell} ,  2{\ell} +1, \rho)=\E^{\rho/2}{\rho}^{-{\ell} -1/2}M_{1/2, {\ell} }(\rho), 
\\ &  
\label{la3.15} 
 Q_{2, 0}(\rho)  =  -\Phi({\ell} +1,  2{\ell} +1, \rho)=-\E^{\rho/2}{\rho}^{-{\ell} -1/2}M_{-1/2, {\ell} }(\rho).
 \end{align}
Now, we need an integral representation from \cite[Section 6.11]{la[1]}:
\begin{equation} \label{la3.16} 
 M_{{\pm}1/2, {\ell} }(\rho)=C({\ell} )(\rho)^{{\ell} +1/2}I_{\pm}(\rho, {\ell} ), \quad C({\ell} )=2^{-{\ell} }\frac{\Gamma(2{\ell} +1)}{\Gamma({\ell} +1)\Gamma({\ell} )},
\end{equation}
where
\begin{equation} \nn
I_{+}(\rho, {\ell} )=\int_{-1}^{1}\E^{t\rho/2}(1-t^{2})^{{\ell} -1}(1-t) d t, \quad I_{-}(\rho, {\ell} )=\int_{-1}^{1}\E^{t\rho/2}(1-t^{2})^{{\ell} -1}(1+t) d t.
\end{equation}
Integrating  the expressions  above  by parts ($\ell$ times),
we obtain: 
\begin{align} &
\label{la3.20}  
I_{+}(\rho, {\ell} )  {\sim}  (-1)^{{\ell} }(\rho/2)^{-{\ell} }\Gamma({\ell} )\E^{-\rho/2}, 
\quad  \rho {\to}\infty \quad (\arg(\rho)=\pi/2),  \\ &
\label{la3.21} 
 I_{-}(\rho,  {\ell} )  {\sim}  -(\rho/2)^{-{\ell} }\Gamma({\ell} )\E^{\rho/2}, 
\quad  \rho {\to}\infty \quad (\arg(\rho)=\pi/2), 
\end{align}
Using asymptotic formulas   \eqref{la3.20} and \eqref{la3.21} we have
\begin{eqnarray} \label{la3.22}  
Q_{1, 0}(\rho) & {\sim} & 
(-1)^{{\ell} }{\rho}^{-{\ell} }\Gamma(2{\ell} +1)/\Gamma({\ell} +1), \quad  
\rho {\to}\infty \quad (\arg(\rho)=\pi/2), \\
\label{la3.23}  
Q_{2, 0}(\rho) & {\sim} & 
-{\rho}^{-{\ell} }\E^{\rho}\Gamma(2{\ell} +1)/\Gamma({\ell} +1), \quad  
\rho {\to}\infty \quad (\arg(\rho)=\pi/2).
\end{eqnarray}
Formulas \eqref{la3.22} and \eqref{la3.23} show that the solutions of \eqref{la3.12} and \eqref{la3.13} of the form  \eqref{la3.14} and \eqref{la3.15}
are, indeed, regular at $\rho =0$ and, moreover, we have
\begin{equation} \label{la3.24}  
Q_{1, 0}{\sim}1, \quad  Q_{2, 0}{\sim}1 \quad {\mathrm{for}} \quad \rho{\to}0.
\end{equation}
The functions
\begin{eqnarray} 
\label{la3.25}  
&& \wt Q_{1, 0}(\rho)=c_{1}\Psi({\ell} , 2{\ell} +1, \rho)=
c_{1}\E^{\rho/2}{\rho}^{-{\ell} -1/2}W_{1/2, {\ell} }(\rho) \\ 
\label{la3.26} 
&& \wt Q_{2, 0}(\rho)=c_{2}\Psi({\ell} +1, 2{\ell} +1, \rho)=
c_{2}\E^{\rho/2}{\rho}^{-{\ell} -1/2}W_{-1/2, {\ell} }(\rho), 
\end{eqnarray}
where $\Psi(a, c, x)$ is the confluent hypergeometric function of the second kind,
also satisfy  \eqref{la3.12} and \eqref{la3.13} but are non-regular at $\rho=0$ (see \cite[Ch.6]{la[1]}).
According to  \cite[Section 6.11]{la[1]},  we have the following integral representation for $W_{{\pm}1/2, {\ell} }$:
\begin{equation} \label{la3.27}  
W_{{\pm}1/2, {\ell} }(\rho)=C_{\pm}({\ell} )\E^{-\rho/2}(\rho)^{{\ell} +1/2}J_{\pm}(\rho, {\ell} ), \quad  C_{\pm}({\ell} )=1/{\Gamma({\ell} +1/2\mp{1/2})}, 
\end{equation}
where
\begin{equation}  \label{la3.30}  
 J_{+}(\rho, {\ell} )=\int_{0}^{\infty }\E^{-t\rho}t^{{\ell} -1}(1+t)^{{\ell} } d t, \quad J_{-}(\rho, {\ell} )=\int_{0}^{\infty }\E^{-t\rho}t^{{\ell} }(1+t)^{{\ell} -1} d t.
\end{equation}
Using  integral representation \eqref{la2.14}, we rewrite  \eqref{la3.30} as
\begin{equation} \label{la3.31}
  J_{+}(\rho, {\ell} )=\rho^{-2{\ell} }
\sum_{i=0}^{{\ell} }C_{{\ell} }^{i}\Gamma(2{\ell} -i)\rho^{i}, 
\quad  
J_{-}(\rho, {\ell} )=\rho^{-2{\ell} }
\sum_{i=0}^{{\ell} -1}C_{{\ell} -1}^{i}\Gamma(2{\ell} -i)\rho^{i}, 
\end{equation}
where $\, C_{{\ell} }^{i} \,$ are the binomial coefficients again. The right-hand sides in \eqref{la3.31} 
are  analytic functions (for $\rho{\ne}0$).  Thus 
\eqref{la3.31} is valid for all $\rho{\ne}0$ in the complex plane.

In view of \eqref{la3.25}--\eqref{la3.27} and \eqref{la3.31}, if $\rho{\to}0$ we have
\begin{equation} \label{la3.33}
  \wt Q_{1, 0}(\rho){\sim}c_1{\rho}^{-2{\ell} }\Gamma(2{\ell} )/\Gamma({\ell} ), \quad  
\wt Q_{2, 0}(\rho){\sim}c_2{\rho}^{-2{\ell} }\Gamma(2{\ell} )/\Gamma({\ell} +1).
\end{equation}
After the substitution $Q_{1,0}=\wt Q_{1, 0}$ and $Q_{2,0}=\wt Q_{2, 0}$ into \eqref{la3.11'}, 
relations \eqref{la3.10} and \eqref{la3.33} imply that we can assume
\begin{equation} \label{la3.34}
   c_{1}=1, \quad  c_{2}={\ell} .
\end{equation}
Using again formulas  \eqref{la3.25}--\eqref{la3.27}
 and \eqref{la3.31}, we derive
\begin{equation} \label{la3.35}
  \wt Q_{1, 0}(\rho){\sim}
{\rho}^{-{\ell} }, \quad  
\wt Q_{2, 0}(\rho){\sim}
{\ell} {\rho}^{-({\ell} +1)}, \quad  \rho {\to}\infty \quad (\arg(\rho)=\pi/2).
\end{equation}
The constructed regular and non-regular (at $r=0$) solutions of \eqref{la1.2}--\eqref{la1.3} (or, equivalently \eqref{la1.9}), where $q\equiv 0$, we denote
by  
\begin{align} &\label{lad1}
F_{0}(r)=F_0(r, \ve, {\ell})=\col [F_{1,0}(r, \ve, \ell) \quad F_{2,0}(r, \ve, {\ell})] 
\end{align}
and
\begin{align}
&\label{lad2}
\wt F_{0}(r)=\wt F_0(r, \ve, {\ell})=\col [\wt F_{1,0} (r, \ve, {\ell}) \quad \wt F_{2,0} (r, \ve, {\ell})], 
\end{align}
respectively. 
According to \eqref{la3.2} and \eqref{la3.11'} we have
\begin{align} &\label{lad3}
  F_{1,0}(r)=\sqrt{m+{z}}\, \E^{-r\ve}(2r\ve)^{{\ell} -1}\big(Q_{1, 0}(r)+Q_{2, 0}(r)\big)r, \\ & \label{lad4}
  F_{2,0}(r)=-\sqrt{m-{z}}\, \E^{-r\ve}(2r\ve)^{{\ell} -1}\big(Q_{1, 0}(r)-Q_{2, 0}(r)\big)r;  
  \\ &
\label{lad5}
 \wt F_{1,0}(r)=\sqrt{m+{z}}\, \E^{-r\ve}(2r\ve)^{{\ell} -1}\big(\wt Q_{1, 0}(r)+\wt Q_{2, 0}(r)\big)r
,
\\ \label{lad6} &
\wt F_{2,0}(r)=-\sqrt{m-{z}}\, \E^{-r\ve}(2r\ve)^{{\ell} -1}\big(\wt Q_{1, 0}(r)-\wt Q_{2, 0}(r)\big)r.
\end{align}
Hence, from  \eqref{la3.22} and  \eqref{la3.23} we obtain
\begin{align}\label{la3.37}
 F_{0}(r, \ve, {\ell}) {\sim}&\frac{\Gamma(2{\ell} +1)}{2\ve\Gamma({\ell} +1)}
\\ &\nn
\times \col \left[ \big((-1)^{{\ell} }\E^{-r\ve}-\E^{r\ve}\big)\sqrt{m+{z}} \quad \,\,
  -\big((-1)^{{\ell} }\E^{-r\ve}+\E^{r\ve}\big)\sqrt{m-{z}}\right] , 
  \end{align}
where ${z}>m$ (i.e., $\arg(\ve(z))=\pi/2$) and either $\ve \to \infty$ or $r \to \infty$.
In a similar way, from \eqref{la3.35} we derive that
\begin{equation} \label{la3.38}
\wt  F_{0}(r, \ve,  {\ell} ){\sim}\frac{1}{2\ve}\E^{-r\ve}
\col \left[ \sqrt{m+{z}} \quad \quad -\sqrt{m-{z}}\right] ,  \quad  {z}>m,
\end{equation}
where either $\ve \to \infty$ or $r \to \infty$

{\bf 2.} Now, we consider the case when $q(r)\not\equiv 0$,  and introduce $2\times 2$ matrix functions
\begin{equation} \label{la3.40}  
\wh V(r)=\left[\begin{array}{cc}
                             0 & q(r) \\
                             -q(r) & 0
                           \end{array}\right], \quad
U_0(r, \ve, {\ell})=\left[
                         \begin{array}{cc}
                         F_{1,0}(r, \ve, {\ell}) & \wt F_{1,0}(r, \ve, {\ell} ) \\
                         F_{2,0}(r, \ve, {\ell} ) & \wt F_{2,0}(r,\ve, {\ell} ) 
                         \end{array}
                       \right]. 
                       \end{equation}
It is easy to see that the  solution  $\wt F(r, \ve, {\ell} )$ of the  equation
\begin{equation} \label{la3.39}  
\wt F(r)=\wt F_{0}(r)+\int_{r}^{\infty }U_0(r)U_0(t)^{-1}\wh V(t)
\wt F(t) d t
\end{equation} 
satisfies the differential system \eqref{la1.2}, \eqref{la1.3}.

The following equality is valid:
\begin{equation} \label{la3.41} 
  \det U_0(r,\ve, \ell)={\Gamma(2{\ell} )}/\big({\ve\Gamma({\ell} )}\big).
\end{equation}
Indeed, it follows from \eqref{la1.2} and  \eqref{la1.3} that
$\big(\det U_0(r, \ve, {\ell} )\big)^{\prime}=0.$ Hence,  we obtain the relation $\det U_0(r, \ve, {\ell} )\equiv \const.$
  Since the first and second columns  of $U_0$ are $F_0$ and $\wt F_0$, respectively,  formulas \eqref{la3.37}--\eqref{la3.40}
 imply that
\begin{equation} \label{la3.42}  \det U_0(r, \ve, {\ell} ){\sim}{\Gamma(2{\ell} )}/\big({\ve\Gamma({\ell} )}\big), 
 \quad  r{\to}\infty  \quad  ({z}>m),
\end{equation}
which yields \eqref{la3.41}.

According to \eqref{la3.37}--\eqref{la3.40} and \eqref{la3.41}, the equality
\begin{align} \label{la3.43}&  
U_0(r)U_0(t)^{-1}=\E^{(r-t)\ve}\Theta_{1}+
\E^{(t-r)\ve}\Theta_{2}
+o(1),   \quad
\ve{\to}\infty   \quad ({z}>m), \\
\label{la3.44}  &
\Theta_1:=\frac{1}{2}
\left[\begin{array}{cc}
  1 & \I \\
  -\I & 1
\end{array}\right], \quad  \Theta_2:=\frac{1}{2}
\left[\begin{array}{cc}
  1 & -\I  \\
  \I & 1
\end{array}\right]
\end{align}
holds.
We multiply both sides of \eqref{la3.39} by  $-2\sqrt{m+{z}}\, \E^{r\ve}$.   Using \eqref{la3.43} and passing to the limit 
${z}{\to}+\infty $,  we obtain
\begin{equation} \label{la3.45}  
\wt F_{\infty }(r, {\ell} )=\begin{bmatrix} \I \\ 1\end{bmatrix}+\int_{r}^{\infty }T_2 \wh V (t)\wt F_{\infty }(t, {\ell} ) d t, 
\end{equation}
where
\begin{equation} \label{la3.46}  
\wt F_{\infty }(r, {\ell} )=-2\lim_{{z}\to{+\infty }}\left(\sqrt{m+{z}}\,
\E^{r\ve}\wt F(r, \ve, {\ell} )\right) .
\end{equation}
The equality
\begin{equation} \label{la3.47}   
\wt F_{\infty }(r, {\ell} )=\E^{-\I \int_{r}^{\infty }q(t) d t} \begin{bmatrix} \I \\ 1\end{bmatrix}
\end{equation}
follows directly from \eqref{la3.38}, \eqref{la3.40}, \eqref{la3.44} and \eqref{la3.45}.
From \eqref{la3.46} and \eqref{la3.47} we see that the constructed solution $\wt F(r,\ve, \ell)$ has the properties stated in Theorem \ref{latm1.2},
that is, Theorem \ref{latm1.2} is proved.

\subsection{Proofs of Theorems   \ref{latm1.7} and \ref{latm1.14}}
\label{lasec4}
In order to show that the results on multidimensional Schr\"odinger equation  from
Subsection \ref{lasub1.1} hold, we should prove Theorems   \ref{latm1.7} and \ref{latm1.14}.
First, we prove the   assertion below (see \cite{la[10], la[13]}).
\begin{La}\label{lala4.1}
The following asymptotic relation
\begin{equation} \label{la4.1}  
\sum_{n=1}^{\infty }\E^{-n^{2}/{z}}=\frac{1}{2}\sqrt{{z}\pi}-
\frac{1}{2}+o(1), \quad  {z}{\to}\infty  
\end{equation} 
is valid.
\end{La}

\begin{proof}
In the proof of  \eqref{la4.1} we use the Poisson formula (see, e.g., \cite{la[4]}):
\begin{equation} \label{la4.2}  
\sum_{n=1}^{\infty }\E^{-n^{2}/{z}}=-\frac{1}{2}+
\int_{0}^{\infty }\E^{-x^{2}/{z}} d x+
2\sum_{n=1}^{\infty }\int_{0}^{\infty }\E^{-x^{2}/{z}}\cos{(2{\pi}nx)} d x.
\end{equation}
Since
\begin{equation} \label{la4.3}  
\int_{0}^{\infty }\E^{-x^{2}/{z}}\cos{(2{\pi}nx)} d x=
\frac{1}{2}\sqrt{{z}\pi}\, \E^{-{{z}}n^{2}\pi^{2}},
\end{equation}
 formula \eqref{la4.2} implies  that
\begin{equation} \label{la4.4}  
\sum_{n=1}^{\infty }\E^{-n^{2}/{z}}=-\frac{1}{2}+\frac{1}{2}\sqrt{{z}\pi}+
\sqrt{{z}\pi}\sum_{n=1}^{\infty }\E^{-{{z}}n^{2}\pi^{2}}.
\end{equation}
Relation \eqref{la4.1} follows directly from \eqref{la4.4}.  
\end{proof}
\textit{Proof of Theorem \ref{latm1.7}}.
In view of \eqref{la1.17} we have
\begin{equation} \label{la4.5}  
{z}_{n}=(n\pi/a)^{2}+(1/a)\delta(a)+o(1), \quad  n{\to}\infty .
\end{equation}
Substituting ${z}=T(a/\pi)^{2}$ into \eqref{la4.1} and taking into account \eqref{la4.5} we derive \eqref{la1.19},
where $Z$ is given by \eqref{la1.18}. $\hspace{15em} \Box$

In the proof of Theorem \ref{latm1.14},
the  next well-known relations 
\begin{align} \label{la4.7}  &
\sum_{n=1}^{\infty }\E^{-2n/T}=\frac{1}{2\sinh (1/T)}=
\frac{1}{2}\left( T+O\left( 1/T\right) \right) , \quad  T{\to}\infty ;
\\ 
 \label{la4.8}  &
\sum_{n=1}^{\infty }\E^{-2n/T}/\sqrt{2n}=
\int_{0}^{\infty }(\E^{-2x/T}/\sqrt{2x}) d x+O(1), \quad  T{\to}\infty
\end{align}
are used instead of the Poisson formula, which was applied in the previous proof.
Since
\begin{equation} \label{la4.9}  
\int_{0}^{\infty }\left( \E^{-u}/\sqrt{u}\right)  d u=\sqrt{\pi} ,
\end{equation}
the assertion of the Theorem \ref{latm1.14} follows from \eqref{la4.7} and \eqref{la4.8}.

\subsection{Dirac system on a finite interval}
\label{lasec5}
The considerations from the previous subsections can be used also to deal with Dirac system on a finite interval.
In this final subsection of the section we consider the Dirac system \eqref{la1.2},  \eqref{la1.3} on the segment $[0, a]$ and assume that
the inequality
\begin{equation} \label{la5.1}   
\int_{0}^{a}|q(t)| d t<\infty 
\end{equation}
holds.
The  solution  $F(r, \ve, \ell)$ of the integral equation
\begin{equation} \label{la5.2}  
F(r,  \ve, \ell)=F_{0}(r, \ve, \ell)-\int_{0}^{r}U_0(r)U_0(t)^{-1}\wh V(t) F(t,  \ve, \ell) d t,
\end{equation} 
where $U_0$ and $\wh V$ are given by  \eqref{la3.40}, satisfies the  system \eqref{la1.2}, \eqref{la1.3}.
Here $F_0$ is a regular solution of the  system \eqref{la1.2}, \eqref{la1.3} with $q\equiv 0$ and is given by
\eqref{lad1}, \eqref{lad3} and \eqref{lad4}.
The parameter $\ve=\ve(z)$ above is the same as in Remark \ref{Rkve} and in Subsection \ref{lasec3}.
Recall that $B(0,\, l]$ stands for the class of  bounded functions on $(0,\, l]$, that is, such functions, the values of which are uniformly bounded 
in the norm. Below we consider functions bounded with respect to the variable $r$.
From formulas  \eqref{la3.14}, \eqref{la3.15}, \eqref{lad3} and \eqref{lad4} we derive
\begin{equation} \label{la5.3}  
F_{0}(r, \ve, \ell)/\left(\sqrt{m+{z}}(2r\ve)^{\ell-1}r\right)\in B(0, \, 1/|\ve|] \quad  ({z}>m).
\end{equation}
In fact the left-hand side of \eqref{la5.3} is bounded when $z>m$ and $\rho =2r\ve$ is bounded.
In a similar way, 
using \eqref{la3.25}, \eqref{la3.26}, \eqref{lad5} and \eqref{lad6}, we have
\begin{equation} \label{la5.4}  
 \wt F_{0} (r, \ve, \ell)/\left(\sqrt{m+{z}}(2r\ve)^{-\ell-1}r\right)\in B(0,\, 1/|\ve|] \quad  ({z}>m). 
\end{equation}
According to \eqref{la3.22}, \eqref{la3.23} and \eqref{la3.33} the relation
\begin{equation} \label{la5.8} 
 U_0(r) \in B(0,\, 1/|\ve|] \quad  ({z}>m)  
\end{equation} is valid.
Relations \eqref{la3.40}, \eqref{la5.3}, \eqref{la5.4} and \eqref{la5.8} imply that
\begin{equation} \label{la5.5}  
 \sqrt{m+{z}}(2r\ve)^{\ell-1}rU_0(r)^{-1} \in B(0,\, 1/|\ve|] \quad  ({z}>m)
\end{equation} 
and
\begin{equation} \label{la5.6} 
\sup\left\| (t/r)^{2\ell} U_0(r)U_0(t)^{-1}\right\|<\infty,   \quad 0\leq t \leq r\leq 1/|\ve| \quad ({z}>m).
\end{equation}
It follows from \eqref{la5.2}, \eqref{la5.3} and \eqref{la5.6} that
\begin{equation} \label{la5.7} 
 F (r, \ve, \ell)/\left(\sqrt{m+{z}}(2r\ve)^{\ell-1}r\right) \in [B(0,\, 1/|\ve] \quad  ({z}>m). 
\end{equation}
In view of \eqref{la5.8}, \eqref{la5.5} and \eqref{la5.7}  the relation
\begin{equation} \label{la5.9} 
 \left\| \int_{0}^{1/|\ve|}U_0(r)U_0(t)^{-1}\wh V(t)
F(t, \ve, \ell) d t\right\| =o(1), \quad \ve \to \infty  \quad ({z}>m) 
\end{equation}
is valid.
Substituting  \eqref{la3.22} and \eqref{la3.23} into \eqref{lad3} and \eqref{lad4}, we represent   $F_0(r, \ve, {\ell})$
(for the case that $\ve=\ve(z)$, $z>m$)  in the form
\begin{align} \label{la5.10} &
 \E^{-\I {\ell}\pi/2}F_0(r, \ve, {\ell})=\E^{-r\ve}\E^{\I{\ell}\pi/2}g_1+\E^{r\ve}\E^{-\I {\ell}\pi/2}g_2+o(1), \quad \ve \to \infty,
\\ & \label{la5.11}
 g_1:=-
\frac{\Gamma(2{\ell}+1)}{2\Gamma({\ell}+1)}\E^{\I{\ell}\pi/2}\left[\begin{array}{c}
                                                                             \I \\
                                                                             1
                                                                         \end{array}\right], 
\quad
 g_2:=
\frac{\Gamma(2{\ell}+1)}{2\Gamma({\ell}+1)}\E^{\I{\ell}\pi/2}\left[\begin{array}{c}
                                                                             \I \\
                                                                             -1
                                                                         \end{array}\right].
\end{align}
Now, we can prove our next theorem.
\begin{Tm}\label{latm5.1}
Let condition \eqref{la5.1} hold.
Then $($for the case that $\ve=\ve(z)$, $z>m)$  we have
\begin{align} \label{la5.13}  &
\E^{-\I {\ell}\pi/2}F(r, \ve, {\ell})=\E^{-r\ve}\E^{\I {\ell}\pi/2}h_1(r)+\E^{r\ve}\E^{-\I {\ell}\pi/2}h_2(r)+o(1), \quad \ve \to \infty,
\\ &
\label{la5.14}  
h_1(r):=-
\frac{\Gamma(2{\ell}+1)}{2\Gamma({\ell}+1)}\E^{\I\delta(r)}\left[\begin{array}{c}
                                                                             \I\\
                                                                             1
                                    \end{array}\right],
\quad
  h_2(r):=
\frac{\Gamma(2{\ell}+1)}{2\Gamma({\ell}+1)}\E^{-\I \delta(r)}\left[
\begin{array}{c}
\I\\
  -1
\end{array}\right],
\end{align}
where
the quantum defect $\delta(r)$has the form
\begin{equation} \label{la5.16}
  \delta(r)=\int_{0}^{r}q(t) d t.
\end{equation}
\end{Tm}

\begin{proof} 
In order to prove the theorem, we represent $F(r, \ve, {\ell})$ in the form
\begin{equation} \label{la5.17}  
\E^{-\I {\ell}\pi/2}F(r, \ve, {\ell})=\E^{-r\ve}\E^{\I {\ell}\pi/2}h_1(r)+\E^{r\ve}\E^{-\I {\ell}\pi/2}h_2(r)+
\wh f(r, \ve, {\ell})
\end{equation} 
and estimate $\wh f$. Definitions \eqref{la5.11} and \eqref{la5.14} imply that
\begin{equation} \label{la5.18}
  h_1(r)=g_1-\int_{0}^{r}T_1\wh V(t)h_1(t) d t, 
\quad
  h_2(r)=g_2-\int_{0}^{r}T_2\wh V(t)h_2(t) d t.
\end{equation}
Let us multiply both sides of \eqref{la5.17} by $\E^{\I \ell \pi/2}$ and substitute the result into  \eqref{la5.2}.
Then, formula  \eqref{la5.18} implies that
the function
\begin{equation} \label{la5.20}  
G(r,  \ve, \ell)=\wh f(r, \ve,{\ell})+\int_{0}^{r}U_0(r)U_0(t)^{-1}\wh V(t)
\wh f(t, \ve,{\ell}) d t
\end{equation} 
satisfies the relation
\begin{equation} \label{la5.21}
  \| G(r, \ve, {\ell})\| {\to}0, \quad  {z}{\to}+\infty .
\end{equation}
Using  \eqref{la5.1}, \eqref{la5.20} and \eqref{la5.21}, we obtain
\begin{equation} \label{la5.22}
  \|\wh f(r, \ve,{\ell})\| {\to}0, \quad  {z}{\to}+\infty .
\end{equation}
\end{proof}
Recall that $F=\col\begin{bmatrix} F_1 & F_2 \end{bmatrix}$ is the regular solution
of the radial Dirac system. Consider the case of the second boundary condition
\begin{equation} \label{la5.23}  
F(a, \ve_{n}, {\ell})\sin {\psi}+F_2(a, \ve_{n}, {\ell})\cos{\psi}=0, \quad  -\pi/2{\leq}\psi{\leq}\pi/2.
\end{equation} 
Here $\ve_{n}=\ve(z_n)=\sqrt{m^{2}-{z}_{n}^{2}}$ and (differently from other considerations of this section, where $n$
always belongs $\BN$) $\,\,n\in \BZ-\{0\}$. Without loss of generality we assume that $z_n>m$
for $n>0$ and $z_n<-m$ for $n<0$.
\begin{Cy}\label{lacy5.2}
Let conditions \eqref{la5.1} and \eqref{la5.23}  be fulfilled.
Then we have
\begin{equation} \label{la5.24}  
{z}_{n}(a)=\frac{\pi}{a}\left(n+\frac{\ell}{2}\right)
+\frac{\psi-\pi/2+\delta(a)}{a}
+o(1).
\end{equation}
\end{Cy}
\begin{proof} 
It follows from Theorem \ref{latm5.1} that
\begin{align} \label{la5.25}  &
F_1(r, \ve_{n}, {\ell})=-\E^{\I {\ell}\pi/2}\frac{\Gamma(2{\ell}+1)}
{\Gamma({\ell}+1)}\sin{\big(rs_n-{\ell}\pi/2-\delta(r)\big)}
+o(1), 
\\ & \label{la5.26}  
F_2(r, \ve_{n}, {\ell})=-\E^{\I {\ell}\pi/2}\frac{\Gamma(2{\ell}+1)}
{\Gamma({\ell}+1)}\cos{\big(rs_n-{\ell}\pi/2-\delta(r) \big)}
+o(1), 
\end{align}
where $\ve_{n}=\I \sqrt{{z}_{n}^{2}-m^{2}}=\I s_n$ $(n>0, \, z_n>m, \, s_n>0)$. In view of \eqref{la5.23}, \eqref{la5.25} and \eqref{la5.26}, the relation
\begin{equation} \label{la5.27}  
\cos{\big(as_n-{\ell}\pi/2-\delta(a)-\psi \big)}=o(1) 
\end{equation} 
is valid.
 Hence the equality \eqref{la5.24} is proved for $n>0$. The case $n<0$
 can be dealt with in the same way.
\end{proof}

 \begin{Rk}\label{lark5.4}
 Formula \eqref{la5.24} is essential  for solving the corresponding inverse sliding problem.
 \end{Rk}

 \begin{Rk}\label{lark5.3}
When $\ell=0$, formula \eqref{la5.24} is well-known  (see \cite[Ch.VII]{LS}).
 \end{Rk}
 
%%%%%%%%%%%%%%%%%%%%%%%%%%%%%%%%% 
%%%%%%%%%%%%%%%%%%%%%%%%%%%% 
%%%%%%%%%%%%%%%%%%%%%%
\section{Schr\"odinger and Dirac equations \\ with Coulomb-type potentials}
\label{LASec2}

\setcounter{equation}{0}

 The radial Schr\"odinger equation with the Coulomb-type potential has the form
\begin{equation} \label{2la1.1}
\frac{d^{2}y}{dr^{2}}+\left({z}+\frac{2a}{r}-\frac{\ell(\ell+1)}{r^{2}}-q(r)\right) y=0, \quad  
0{\leq}r<\infty, \quad  a=\overline{a}\not= 0,  
\end{equation}
where $\ell=0, 1, 2, \dots$ Its potential differs from the potential in \eqref{la1.1} by the additional term $\frac{2a}{r}$.
The radial  Dirac system (relativistic case) with the Coulomb-type potential has the form:
\begin{align} \label{2la1.2}&
\left(\frac{d}{dr}+\frac{\ell}{r}\right) f_1-\left({z}+m+\frac{a}{r}-q(r)\right) f_2=0,
\\  \label{2la1.3} &
   \left(\frac{d}{dr}-\frac{\ell}{r}\right) f_2+\left( {z}-m+\frac{a}{r}-q(r)\right) f_1=0, 
\\  \label{2la1.3'} &   
   0{\leq}r<\infty, \quad m>0, \quad q=\ov{q}, \quad    a=\overline{a}\not= 0,  \quad \ell=\pm 1, \pm 2, \dots, \quad \ell^2 >a^2.
\end{align}
Equations \eqref{2la1.2} and  \eqref{2la1.3} differ from the equations \eqref{la1.2} and  \eqref{la1.3}
by the additional term $\frac{a}{r}$.
Like in Section \ref{LASec1}, without loss of generality
we consider only the case $\ell>0$. 

The scheme presented in Section \ref{LASec1} admits modification for the case of Coulomb-type potentials.
Equation \eqref{2la1.1} and system \eqref{2la1.1}, \eqref{2la1.2} are essential in the study of the spectrum of atoms and molecules
\cite{la[2], la[3], 2la[4]}.

We assume that
\begin{equation} \label{2la1.4}   
\int_{r}^{\infty}|q(t)|dt<\infty, \quad  r>0.
\end{equation}
In the present section we describe the asymptotic behavior of the solutions of equation
\eqref{2la1.1} and 
of system \eqref{2la1.2}, \eqref{2la1.3} for the energy tending to infinity
 (i.e., ${z}{\to}\infty$). 
 Using this asymptotics, we  introduce the notion of the quantum defect and solve the corresponding sliding inverse problem
 for the relativistic case.

 \subsection{Asymptotics of the solutions:  Schr\"odinger equation }
 \label{2lasec2}
 If $q(r)\equiv 0$, then  \eqref{2la1.1} takes the form:
\begin{equation} \label{2la2.1}   
\frac{d^{2}y}{dr^{2}}+\left( {z}+\frac{2a}{r}-\frac{\ell(\ell+1)}{r^{2}}\right) y=0.
\end{equation}
We denote by $u_1$ and $u_{2}$ the  solutions of  \eqref{2la2.1}
such that
\begin{equation} \label{2la2.2}  
 u_{1}(r, \ve, \ell)=(2\ve)^{-(\ell+1)}M_{a/\ve, \ell+1/2}(2r\ve), \quad
u_{2}(r, \ve, \ell)=(2\ve)^{\ell}W_{a/\ve, \ell+1/2}(2r\ve), 
\end{equation}
 where $M_{\kappa, \mu}(z)$ and $W_{\kappa, \mu}(z)$ are   Whittaker functions.
Like in Subsection \ref{lasec2}, the parameter $\ve=\I \sqrt{{z}}$ is defined for all ${z}$  in the complex 
plane with the cut along the negative  part of the imaginary  axis.
We put $\arg{\ve}={\pi}/2$ for ${z}>0$.
Recall the connections between the  Whittaker  and
confluent hypergeometric functions (see \cite{la[1]}):
\begin{equation} \label{2la2.3}
   M_{a/\ve, \ell+1/2}(x)=\E^{-x/2}x^{c/2}\Phi(\a, c, x), \quad  
W_{a/\ve, \ell+1/2}(x)=\E^{-x/2}x^{c/2}\Psi(\a, c, x), \end{equation}
where $   \a=\ell+1-a/\ve, \quad  c=2\ell+2$.
Instead of asymptotics in \eqref{la2.4}, we need now the asymptotics of  $\Phi$ and $\Psi$ for energy tending to infinity.
More precisely, we need the following well-known  formulas
(see \cite[Section 6.13]{la[1]}):
\begin{align} \label{2la2.6}&
   \Phi(\a, c, 2r\ve){\sim}\frac{\Gamma(2\ell+2)}{\Gamma(\ell+1)}(-2r\ve)^{-\ell-1}+
\frac{\Gamma(2\ell+2)}{\Gamma(\ell+1)}
(2r\ve)^{-\ell-1}\E^{2r\ve}, 
\\ &
 \label{2la2.5}   
\Psi(\a, c, 2r\ve){\sim}(2r\ve)^{-\ell-1}, \,\,  {\mathrm{where}} \,\, \ve{\to}\infty \,\,  {\mathrm{and}} \,\, \arg(\ve(z))= \pi/2 \,\, ({\mathrm{i.e.,}} \, z>0). 
\end{align}
Using \eqref{2la2.6} and \eqref{2la2.5}, we obtain
 the relations
\begin{align}&
\label{2la2.7}
  u_1(r, \ve) {\sim}  \frac{\Gamma(2\ell+2)}{\Gamma(\ell+1)}
\left( (-1)^{\ell+1}\E^{-r\ve}+\E^{r\ve}\right) (2\ve)^{-(\ell+1)}, \,\,  \ve{\to}\infty, \,\,  {z}>0, \\
\label{2la2.8}&
 u_2(r, \ve)  {\sim}  (2\ve)^{\ell}\E^{-r\ve}, \quad  \ve{\to}\infty, \quad  {z}>0.
\end{align}

Next, we consider the general-type Schr\"odinger equation \eqref{2la1.1} (where $q$ is not necessarily trivial). The solution $u(r, \ve, \ell)$ of the integral equation
\begin{equation} \label{2la2.9}
   u(r, \ve, \ell)=u_2(r, \ve, \ell)-\int_{r}^{\infty}
k(r, t, \ve, \ell)q(t)u(t, \ve, \ell)dt,
\end{equation} 
where
the kernel $k(r, t, \ve, \ell)$ is defined by 
\begin{equation} \label{2la2.10}
   k(r, t, \ve, \ell)=\frac{\Gamma(\ell+1-a/\ve)}{\Gamma(2\ell+2)}\big(u_1(r, \ve, \ell)u_2(t, \ve, \ell)-
u_2(r, \ve, \ell)u_1(t, \ve, \ell)\big),
\end{equation}
satisfies  \eqref{2la1.1}.
It follows from \eqref{2la2.7}, \eqref{2la2.8} and \eqref{2la2.10} that
\begin{equation} \label{2la2.11}
   k(r, t, \ve, \ell){\sim}
 \E^{(r-t)\ve}-\E^{-(r-t)\ve} , \quad  \ve{\to}\infty, \quad  {z}>0.
\end{equation}
Using standard methods, we deduce from \eqref{2la2.8}--\eqref{2la2.11}  the following statement.

\begin{Tm}\label{2latm2.1}
Let  \eqref{2la1.4} hold. Then,  for  $ u(r, \ve, \ell)$ satisfying \eqref{2la2.9} we have
\begin{equation} \label{2la2.12}
   u(r, \ve, \ell){\sim}(2\ve)^{-\ell}\E^{-r\ve}, \quad  \ve{\to}\infty, \quad  {z}>0.\end{equation}
\end{Tm}

\subsection{Asymptotics of the solutions: Dirac system}
\label{2lasec3}

If $q(r)\equiv 0$, system \eqref{2la1.2}, \eqref{2la1.3} takes the form
\begin{equation} \label{2la3.1}
   \left(\frac{d}{dr}+\frac{\ell}{r}\right)f_1-({z}+m+\frac{a}{r})f_2=0, \,\,   \left(\frac{d}{dr}-\frac{\ell}{r}\right)f_2+({z}-m+\frac{a}{r})f_1=0 \,\, (\ell^2>a^2).
\end{equation}
We consider  solutions  of  \eqref{2la3.1} in the form similar to \eqref{la3.2} and \eqref{la3.2'}, that is,
\begin{align} \label{2la3.3}  &
f_1(r,\ve, \ell)=\sqrt{m+{z}}\, \E^{-r\ve}(2r\ve)^{{\om} -1}\big(Q_{1}(r,\ve,\ell)+Q_{2}(r,\ve,\ell)\big)r, \\
\label{2la3.3'}  &
f_2(r,\ve,\ell)=-\sqrt{m-{z}} \, \E^{-r\ve}(2r\ve)^{{\om} -1}\big(Q_{1}(r,\ve,\ell)-Q_{2}(r,\ve,\ell)\big)r, 
\end{align}
where   $\om=\sqrt{\ell^{2}-a^{2}}>0$, $\ve=\sqrt{m^{2}-{z}^{2}}$ and the choice of arguments in
$\sqrt{m\pm z}$ and $\sqrt{m^{2}-{z}^{2}}$ is prescribed in Remark \ref{Rkve}.
For regular and non-regular solutions of \eqref{2la3.1} we use here the same notations as
for regular and non-regular solutions of \eqref{la1.2}, \eqref{la1.3}, where $q\equiv 0$, in Section \ref{LASec1}.
For the case of regular (at $r=0$) solutions of \eqref{2la3.1}, the functions $Q_{1}$ and $Q_{2}$  
can be expressed via the confluent hypergeometric functions $\Phi(\a, c, x)$ (see \cite{la[2], la[3]}).
Thus, a regular solution $F_0$ of \eqref{2la3.1} is given by
\begin{align} \label{2la3.3!}  &
F_{0}=\col\begin{bmatrix}F_{1,0} & F_{2,0} \end{bmatrix},\\
\label{2la3.3!!}  &
F_{1,0}(r,\ve, \ell)=\sqrt{m+{z}}\, \E^{-r\ve}(2r\ve)^{{\om} -1}\big(Q_{1,0}(r,\ve,\ell)+Q_{2,0}(r,\ve,\ell)\big)r, \\
\label{2la3.3!!!}  &
F_{2,0}(r,\ve,\ell)=-\sqrt{m-{z}} \, \E^{-r\ve}(2r\ve)^{{\om} -1}\big(Q_{1,0}(r,\ve,\ell)-Q_{2,0}(r,\ve,\ell)\big)r, 
\end{align}
where
\begin{align} 
\label{2la3.6}
   Q_{1,0}  =  \a_1\Phi(\om-a{z}/\ve, 2\om+1, 2r\ve), 
\quad
   Q_{2,0}  =  \a_2\Phi(\om+1-a{z}/\ve, 2\om+1, 2r\ve), 
\end{align}
and
\begin{equation} \label{2la3.8}
   \frac{\a_1+\a_2}{\a_1-\a_2}=-\frac{a}{\om+\ell}
\sqrt{\frac{m-{z}}{m+{z}}}.\end{equation}
Using asymptotic formulas (see \cite{la[1]}) for  the confluent hypergeometric functions $\Phi(\a, c, x)$, we obtain
\begin{align} &
\label{2la3.9}
  Q_{1,0}(r,\ve,\ell) {\sim}  \widehat{\a}_1\frac{\Gamma(2\om+1)}{\Gamma(\om+1-\I a)}
(2r\ve)^{-\om-\I a} , \quad  \ve{\to}\infty \quad  ({z}>m), 
 \\ 
\label{2la3.10} &
   Q_{2,0}(r,\ve,\ell)  {\sim}  \widehat{\a}_2\frac{\Gamma(2\om+1)}{\Gamma(\om+1+\I a)}\E^{2r\ve}
(2r\ve)^{-\om+\I a} , \quad  \ve{\to}\infty \quad  ({z}>m), 
\end{align}
where
\begin{equation} \label{2la3.11}   
\frac{\widehat{\a}_1+\widehat{\a}_2}{\widehat{\a}_1-\widehat{\a}_2}=-\frac{\I a} {\om+\ell}.
\end{equation}
A non-regular at $r=0$
solution $\wt F_{0}=\col [\wt F_{1,0} \quad \wt F_{2,0}]$ of \eqref{2la3.1} has the form
(see \cite{la[2], la[3]}):
\begin{align} \label{2la3.12}&
   \wt F_{1,0}(r,\ve, \ell)=\sqrt{m+{z}}\, \E^{-r\ve}(2r\ve)^{\om-1}(\wt Q_{1,0}(r,\ve, \ell)+\wt Q_{2,0}(r,\ve, \ell))r, \\
   \label{2la3.12'}&
\wt F_{2,0}(r,\ve, \ell)=-\sqrt{m-{z}} \, \E^{-r\ve}(2r\ve)^{\om-1}(\wt Q_{1, 0}(r,\ve, \ell)-\wt Q_{2, 0}(r,\ve, \ell))r, 
\end{align}
where
\begin{align} &
\label{2la3.13}
    \wt Q_{1, 0}(r,\ve, \ell)  = \wt \a_{1}\Psi(\om-{az}/\ve, 2\om+1, 2r\ve),\\
&\label{2la3.14}   
\wt Q_{2, 0}(r,\ve, \ell) =  \wt \a_{2}\Psi(\om+1-{az}/\ve, 2\om+1, 2r\ve)
.
\end{align}
Here  $\Psi(a, c, x)$ is the confluent hypergeometric function of the second kind.
In view of \eqref{2la3.13} and \eqref{2la3.14} we have (see \cite[Ch.6]{la[1]}):
\begin{equation} \label{2la3.15}    
\wt Q_{1, 0}{\sim}{\wt \a}_1{(2r\ve)}^{-2\om}, \quad   \wt Q_{2, 0}{\sim}{\wt \a}_2{(2r\ve)}^{-2\om},  \quad r \to \infty;
\quad  {\wt \a}_1=\frac{\ell\ve-am}{\om\ve+a{z}}{\wt \a}_2.\end{equation}
Using again asymptotic formulas for   $\Phi(\a, c, x)$ (see \cite{la[1]}), we obtain
\begin{equation} \label{2la3.16}
    \wt Q_{1, 0}{\sim}\breve{\a}_1{(2r\ve)}^{-\om-\I a} , \quad  \ve{\to}\infty \quad  ({z}>m), \end{equation}
\begin{equation} \label{2la3.17}
    \wt Q_{2, 0}{\sim}\breve{\a}_2{(2r\ve)}^{-\om-\I a-1}, \quad  \ve{\to}\infty \quad  ({z}>m), \end{equation}
where
\begin{equation} \label{2la3.18}
   \breve{\a}_1=\frac{\ell}{\om-\I a} \breve{\a}_2.\end{equation}
From \eqref{2la3.3!!}, \eqref{2la3.3!!!}, \eqref{2la3.9} and \eqref{2la3.10} we obtain
\begin{align}
\nonumber
 F_{1,0}(r, \ve, \ell){\sim}&\sqrt{m+{z}}\, \Gamma(2\om+1)\\
 \label{2la3.20}
 &\times 
\left(\E^{-r\ve}\widehat{\a}_1\frac{(2r\ve)^{-\I a} }{2\ve\Gamma(\om+1-\I a)}+
\E^{r\ve}\widehat{\a}_2\frac{(2r\ve)^{\I a} }{2\ve\Gamma(\om+1+\I a)}\right) , 
\\ \nn
F_{2,0}(r,  \ve, \ell){\sim}&-\sqrt{m-{z}}\, \Gamma(2\om+1)   \\ 
 \label{2la3.21}
  &\times
\left(\E^{-r\ve}\widehat{\a}_1\frac{(2r\ve)^{-\I a} }{2\ve\Gamma(\om+1-\I a)}-
\E^{r\ve}\widehat{\a}_2\frac{(2r\ve)^{\I a} }{2\ve\Gamma(\om+1+\I a)}\right),
\end{align}
where $\ve{\to}\infty$,  ${z}>m.$
Formulas \eqref{2la3.12}, \eqref{2la3.12'} and \eqref{2la3.16}--\eqref{2la3.18} imply that
\begin{equation} \label{2la3.22}
   \wt F_{0}(r,\ve,\ell){\sim}\frac{1}{2\ve}\breve{\a}_1\E^{-r\ve}{(2r\ve)}^{-\I a} 
\col [\sqrt{m+{z}} \quad -\sqrt{m-{z}}], \quad  \ve{\to}\infty \quad  ({z}>m).
\end{equation}

Next, we consider the case $q(r)\not\equiv 0.$
 The  solution  $\wt F(r,  \ve, \ell)$ of
the integral equation
\begin{equation} \label{2la3.23}
   \wt F(r,  \ve,\ell)=\wt F_{0}(r, \ve, \ell)+\int_{r}^{\infty}U_0(r)U_0(t)^{-1}\wh V(t)
\wt F(t,  \ve, \ell)dt,\end{equation} 
where $\wh V$ and $U_0$ have the form \eqref{la3.40}, 
satisfies  system \eqref{2la1.2}, \eqref{2la1.3}.
We note that  $F_{i,0}$ and $\wt F_{i,0}$ $(i=1,2)$ 
in Section \ref{LASec1} are different from the entries $F_{i,0}$ and $\wt F_{i,0}$ of $U_0$, which are introduced in this section.
For the present case,
 formulas \eqref{2la3.20}-\eqref{2la3.22}
 imply that
\begin{equation} \label{2la3.26}   
 \det U_0(r, \ve, \ell){\sim}-\widehat{\a}_2\breve{\a}_1\frac{\Gamma(2\om+1)}{2\ve\Gamma(\om+1+\I a )}, 
 \quad  \ve{\to}\infty \quad  ({z}>m).
\end{equation}
According to  \eqref{2la3.20}--\eqref{2la3.22} and \eqref{2la3.26} the equality \eqref{la3.43} from Section 
\ref{LASec1} is valid again.
Hence, multiplying both sides of \eqref{2la3.23} by  $-2\sqrt{m+{z}}\,\E^{r\ve}{(2r\ve)}^{\I a} $ and   passing to the limit 
${z}{\to}+\infty$,  we obtain
\begin{equation} \label{2la3.29}
   \wt F_{\infty}(r,  \ell)=\begin{bmatrix}\I \\ 1\end{bmatrix}+\int_{r}^{\infty}\Theta_{1}\wh V(t)\wt F_{\infty}(t,  \ell)dt, 
   \end{equation}
where
\begin{equation} \label{2la3.30}   
\wt F_{\infty}(r,  \ell)=-2\lim_{{z}\to{+\infty}}\left( \sqrt{m+{z}}\,
\E^{r\ve}{(2r\ve)}^{\I a} \wt F(r, \ve, \ell)\right) .
\end{equation}
The equality
\begin{equation} \label{2la3.31}   
 \wt F_{\infty}(r, \ell)=\E^{-\I \int_{r}^{\infty}q(t)dt} \begin{bmatrix}\I \\ 1\end{bmatrix}
 \end{equation}
follows directly from  \eqref{2la3.29}-\eqref{2la3.31}.
Thus, we proved that  $\wt F(r, \ve, \ell)$ constructed above satisfies the requirements
of the following statement. 

\begin{Tm}\label{2latm3.2}
Let condition \eqref{2la1.4} be fulfilled. Then there exists a solution $\wt F(r, \ve, \ell)$
of system \eqref{2la1.2}, \eqref{2la1.3}, which satisfies the relation
\begin{equation} \label{2la3.32}
   \wt F(r, \ve, \ell){\sim}-\frac{1}{2\sqrt{m+{z}}}\, 
\E^{-\I (r\ve+\delta(r))}{(2r\ve)}^{-\I a}  \begin{bmatrix}\I \\ 1\end{bmatrix}, \quad  \ve{\to}\infty \quad  ({z}>m),
\end{equation} 
where the quantum defect $\delta(r)$ is given by the formula
\begin{equation} \label{2la3.33}
   \delta(r)=\int_{r}^{\infty}q(t)dt.
\end{equation}
\end{Tm}

Finally, we formulate the sliding half-inverse problem.

\begin{Pb}\label{2laprobl3.3}
Recover the potential $q-a/r$ of the Dirac system \eqref{2la1.2}, \eqref{2la1.3} from the given quantum defect $\delta$ and constant
$a$.
\end{Pb}

According to  \eqref{2la3.33}, the solution of Problem \ref{2laprobl3.3} has the form:
\begin{equation} \label{2la3.34}
   q(r)=-\frac{d}{dr}\delta(r).
   \end{equation}

\end{document}